# High-Speed Demodulation of weak FBGs Based on Microwave Photonics and Chromatic Dispersion

Lei Zhou [1], Zhengying Li[1,2, *], Na Xiang [1]

[1]Key Laboratory of Fiber Optic Sensing Technology and Information Processing (Wuhan University of Technology), Ministry of Education, Wuhan 430070, China
[2] National Engineering Laboratory for Fiber Optic Sensing Technology, Wuhan University of Technology, Wuhan 430070, China
*Corresponding author: zhyli@whut.edu.cn

**A high speed quasi-distributed demodulation method based on the microwave photonics and the chromatic dispersion effect is designed and implemented for weak fiber Bragg gratings (FBGs). Due to the effect of dispersion compensation fiber (DCF), FBG wavelength shift leads to the change of the difference frequency signal at the mixer. With the way of crossing microwave sweep cycle, all wavelengths of cascade FBGs can be high speed obtained by measuring the frequencies change. Moreover, through the introduction of Chirp-Z and Hanning window algorithm, the analysis of difference frequency signal is achieved very well. By adopting the single-peak filter as a reference, the length disturbance of DCF caused by temperature can be also eliminated. Therefore, the accuracy of this novel method is greatly improved, and high speed demodulation of FBGs can easily realize. The feasibility and performance are experimentally demonstrated using 105 FBGs with 0.1% reflectivity, 1 m spatial interval. Results show that each grating can be distinguished well, and the demodulation rate is as high as 40 kHz, the accuracy is about 8 pm.**

**OCIS codes:** *Fiber Bragg grating; microwave photonics; chromatic dispersion; high-speed demodulation.*

http://dx.doi.org/10.1364/OL.99.099999

Fiber Bragg gratings (FBGs) can be made into a variety of sensors, and easy to form a sensing network, which have been widely applied in construction, chemical, military and aviation [1, 2]. But in some special fields such as aero-engine testing [3], vibration detection [4], structural health monitoring [5], the detection of high frequency signal is limited by traditional demodulation techniques. As a result, high speed demodulation is one of the key technologies to solve the practical application of FBGs in special fields.

Recently, weak FBG has attracted a close attention because of its extremely weak reflectivity, which will effectively reduce the crosstalk between multiple gratings, and greatly increase the multiplexing capacity and scale of FBG sensing network [6, 7]. Many scholars have carried out active exploration in high speed demodulation of weak FBGs. By the method of Fourier domain mode locked (FDML) swept laser [8], eighteen weak gratings are interrogated simultaneously with a rate of 120 kHz. However, due to the wavelength sweeping rate of 2.66 nm/μs, the spatial resolution of FBGs is 20 m. A time-division multiplexing (TDM) sensing network with 12 weak FBGs is verified [9], in which spatial resolution is 0.2 m and accuracy is 10 pm. But limited to the establishment time of the tunable laser, it takes 1.5 min to complete once detection. For improving the demodulation speed, this group demonstrated a high speed technique for weak FBGs by a pulsed light source and a chromatic dispersion unit, and demodulated speed of 20 kHz is experimentally realized [10]. But this method requires extremely high sampling rate, such as 40 GHz, which result in difficult engineering application. A demodulation method of weak FBGs with 10.2 cm spatial resolution is also mentioned based on monitoring the microwave responses [11]. But its demodulation speed is limited by the scanning time of vector network analyzer (VNA) and the optical band-pass filter. In addition, the demodulation techniques adopting frequency-shifted interferometry [12], optical low coherent reflection measurement (OLCR) [13], microwave assisted technique [14] and unbalanced Mach-Zehnder interferometer [15] are also reported. However, wavelength scanning or tunable filter still cannot be avoided generally, which induces a slow demodulation speed of FBGs. And some of these schemes also have problems such as poor stability and complex structure.

In this Letter, we propose a high speed quasi-distributed demodulation method based on the microwave photonics and the chromatic dispersion effect. The scheme uses broadband light source and dispersion compensation fiber (DCF), and wavelength shift of weak FBGs are transformed into frequency domain. By introducing the Chirp-Z and Hanning window algorithm, the analysis of difference frequency signal which demodulation based is achieved very well. Thus, with the way of crossing microwave sweep cycle, all wavelengths of cascade FBGs can be high speed demodulated. Since the difference frequency signal produced by the mixer is very low, high sampling rate is not needed in the demodulation. Furthermore, we eliminate the length disturbance of DCF caused by temperature with adopting the single-peak filter as a reference. In contrast to traditional microwave demodulation technologies [16-18], requiring none of wavelength scanning, tunable filter, VNA or high sampling rate oscilloscope, the demodulation of weak FBGs at high speed is easily realized with this scheme. The principles of above methods are described in detail, and experiments are carried out.

The principle diagram is shown in Fig.1. A broadband light from amplified spontaneous emission (ASE) laser is electro-optically modulated with a microwave (MW) signal. At the output of electro-optic modulator (EOM), the optical signal is split into two ways. One is reflected by the FBG sensing network after passing the circulator, and the other is transmitted by the single-peak filter (SPF). The SPF has characteristic that its peak wavelength doesn't change with temperature, which is used as a reference. Then, the two ways of optical signal are coupled into the delay fiber and DCF. For the realization of crossing sweep cycle accurately, a section of delay fiber is adopted. And DCF is use to provide dispersion for the wavelength to frequency conversion. In the system, the optical band-pass filter and Erbium-doped optical fiber amplifiers (EDFA) are used to improve the signal-noise ratio (SNR). After the photodiode, the signals from FBGs and SPF produce difference frequency with the original MW signal at the mixer and finally collected by data acquisition (DAQ).

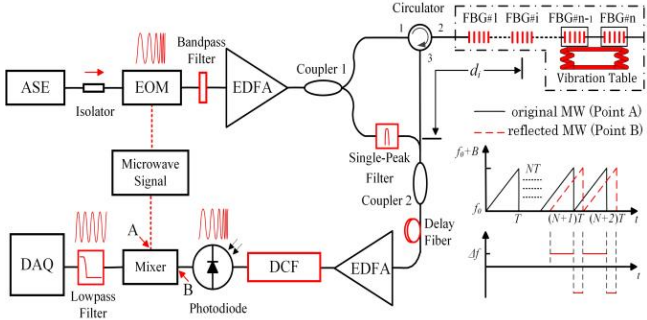

Fig. 1. Schematic of the proposed system based on microwave photonics and chromatic dispersion.

The MW signal whose frequency varies linearly with time can be described as:

$$f(t) = f_0 + Bt/T, 0 \leq t < T \quad (1)$$

Where $B$ and $T$ are, respectively, the frequency bandwidth and the sweep cycle, while $f_0$ and $t$ are the initial frequency and the time variable. In order to acquire the difference frequency signal of the mixer, a microwave low-pass filter is used. After filtering the high frequency by this filter, the difference frequency signal $V(t)$ can be expressed as:

$$V(t) = V \cos(2\pi B t \Delta t / T + \varphi) \quad (2)$$

Therefore, the frequency of $V(t)$ can be described as:

$$\Delta f = B \Delta t / T = B(2n_{\text{eff}} L + n_{\text{DCF}} L_{\text{DCF}})/(Tc) \quad (3)$$

Where $\Delta t$ is the time difference between the original and the reflected MW signal. $V$, $\varphi$, $L$ and $L_{\text{DCF}}$ are the signal amplitude, the initial phase, the length of sensing fiber and the length of DCF, respectively. $n_{\text{eff}}$ and $n_{\text{DCF}}$ are the effective indices of sensing fiber and DCF, $c$ is the speed of the light in vacuum.

By Eq. (3), it is known that the difference frequency signal $\Delta f$ depends on the time difference $\Delta t$. With a long length of DCF providing the dispersion in system, $\Delta t$ will be quite large. In order to avoid the frequency aliasing, a long sweep cycle $T'$ ($T' > \Delta t$) is generally required, as shown in Fig. 2(a). In picture, the original and reflected MW signal are shown with black solid and colorful dash line, respectively. However, long sweep cycle $T'$ means slow demodulation rate. For high speed demodulation, we take the way of crossing sweep cycle to form the difference frequency signal, as see in Fig. 2(b) and Fig. 2(c). The sweep cycle $T$ is set smaller than the time difference $\Delta t$. In this way, the relationship between the sweep cycle $T$ and the length of sensing fiber $L$ should firstly be satisfied:

$$T < 2n_{\text{eff}} L / c \quad (4)$$

Secondly, it should be satisfied:

$$\begin{cases} 0 < (2n_{\text{eff}} L_{\text{SPF}\#r} + n_{\text{DCF}} L_{\text{DCF}})/c - NT < T \\ 0 < (2n_{\text{eff}} L_{\text{FBG}\#n} + n_{\text{DCF}} L_{\text{DCF}})/c - NT < T, (N=0,1,2\cdots) \end{cases} \quad (5)$$

In Eqs. (4) and (5), $L_{\text{SPF}\#r}$ and $L_{\text{FBG}\#n}$ are the position of the single-peak filter and the last grating, respectively. $N$ is the number of cycles. When above relations are satisfied, the single-peak filter and all of FBGs produce difference frequency only with the $N+1$ cycle MW signal, which effectively improved the demodulation speed for $N$ times. Thus, the frequency aliasing will not occur and high speed demodulation is easily to realize.

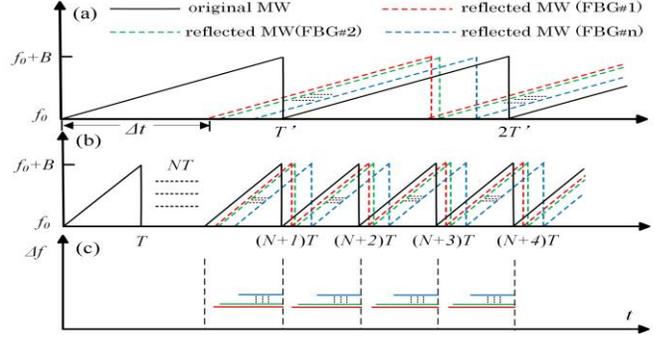

Fig. 2. The forming way of difference frequency signal (a) the generally way, (b) the way of crossing sweep cycle, (c) the frequency of difference frequency signal.

For the purpose of demodulating FBG wavelength more and more accurately, it is necessary to consider the length disturbance of DCF caused by external temperature in long time demodulation. We adopt the single-peak filter as a reference to eliminate it. $\Delta f_{\text{SPF}\#r}$ and $\Delta f_{\text{FBG}\#i}$ are the difference frequency signal produced by the single-peak filter SPF#r and the sensing grating FBG#i respectively, and can be expressed as:

$$\begin{cases} \Delta f_{\text{SPF}\#r} = [2n_{\text{eff}} L_{\text{SPF}\#r}/c + (L_{\text{DCF}} + \Delta L_{\text{DCF}})n_{\text{DCF}}/c - NT]B/T \\ \Delta f_{\text{FBG}\#i} = [2n_{\text{eff}}(L_{\text{SPF}\#r} + d_i)/c + (L_{\text{DCF}} + \Delta L_{\text{DCF}})n_{\text{DCF}}/c \\ \qquad + D_{\text{DCF}}(\lambda_i - \lambda_r) - NT]B/T \end{cases} \quad (6)$$

In Eq. (6), $d_i$ is the distance between SPF#r and FBG#i, and $\Delta L_{\text{DCF}}$ is the length variation of DCF induced by disturbances. $D_{\text{DCF}}(\lambda_i-\lambda_r)$ is the time delay caused by the dispersion effect. Wherein, $\lambda_r$ and $\lambda_i$ are the center wavelengths of SPF#r and FBG#i, respectively, and $D_{\text{DCF}}$ is the total dispersion of DCF.

Through DAQ to collect signals of the photodiode, the frequency difference $f_i$ between $\Delta f_{\text{SPF}\#r}$ and $\Delta f_{\text{FBG}\#i}$ can be obtained with the demodulation algorithm, and can be described as:

$$f_i = [2n_{\text{eff}} d_i / c + D_{\text{DCF}}(\lambda_i - \lambda_r)]B/T \quad (7)$$

As shown in Eq. (7), the length variation $\Delta L_{\text{DCF}}$ is offset by using the single-peak filter as a reference, which makes the stability and accuracy further improved. Meanwhile, due to the single-peak with a constant wavelength $\lambda_r$, when the wavelength $\lambda_i$ of FBG#i changes, it will cause the frequency difference $f_i$ change. This relation can be expressed as:

$$\Delta \lambda_i = \Delta f_i T / (B D_{\text{DCF}}) \quad (8)$$

Wherein, $\Delta \lambda_i$ and $\Delta f_i$ are the variations of $\lambda_i$ and $f_i$, respectively. Thus, the wavelength demodulation of weak FBGs can be realized only by the frequency variation $\Delta f_i$ according to Eqs. (8) obviously.

Fast Fourier transform (FFT) is used for frequency analysis in our demodulation algorithm. However, it is known that the frequency resolution of FFT depends on the sampling time. If $T$ and $B$ are set as 25 μs and 1 GHz, a wavelength resolution of 1 pm requires a frequency resolution of 80 Hz according to Eq. (8). To achieve this resolution should sample 500 cycles. Obviously, the sampling time is too long to achieve high speed demodulation. Therefore, we introduce Chirp-Z algorithm to zoom the frequency signal. One cycle of signals with 20.05 MHz frequency is simulated and moved 1 kHz. We use FFT and Chirp-Z to process it.

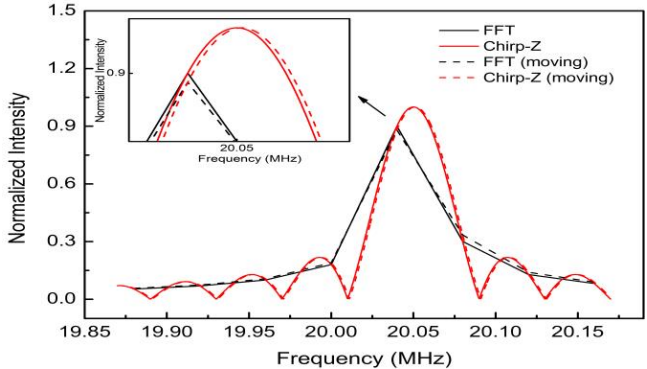

Fig. 3. The frequency spectrum processed by FFT and Chirp-Z.

As shown in Fig. 3, black lines are the processed results of FFT, while red lines for Chirp-Z. The dash lines are the results when frequency moving. For $T$ of one cycle signal is 25 μs, the frequency resolution of FFT is 40 kHz in Fig. 3, which is hard to locate the peak and display the change. By adopting Chirp-Z algorithm, the zooming frequency resolution is 50 Hz, which can accurately locate the peak of 20.05 MHz and display the variation of 1 kHz.

However, when dealing with the difference frequency signals of multi-grating, there is still a problem to be solved. The grating spacing of our FBG network is 1 m, so we simulate the signal of two gratings with 1 m spacing and process it by Chirp-Z algorithm. As shown in Fig. 4(a) with black line, the frequency sidelobes overlap each other and will influence the peak-finding accurate. By using Hanning window to reduce the spectrum leakage, the result is shown in the red line of Fig. 4(a). The spectrum leakage is very small and almost no sidelobes overlap. We analysis the peak-finding accurate of Chirp-Z and Chirp-Z + Hanning algorithms by simulation, and the results are given in Fig. 4(b) and Fig. 4(c). In the simulation, the frequency of FBG#1 is constant while FBG#2 is shifted 2 kHz each time.

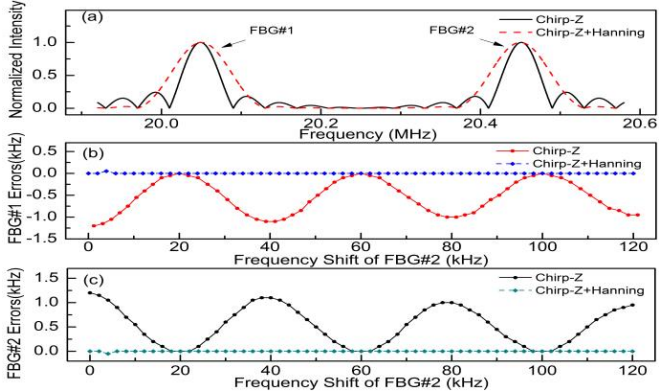

Fig. 4. The processing results of Chirp-Z and Chirp-Z + Hanning algorithm (a) the frequency spectrum of two algorithm, (b) the peak-finding errors of FBG#1, (c) the peak-finding errors of FBG#2.

In Fig. 4(b), when only using the Chirp-Z algorithm, the overlap still has an influence on the accuracy. The frequency of FBG#1 should not change, but the max change of FBG#1 is 1200 Hz, which is about 15 pm of wavelength error due to Eq. (8). And the frequency shift of FBG#2 also cannot match the setting value of horizontal axis. However, almost no influence when using Chirp-Z + Hanning window. The frequency of FBG#1 is not changed and the shift of FBG#2 also match the setting. In Fig. 4(c), there is still an error of FBG#2 when using Chirp-Z without Hanning window. Due to the spectrum energy interacts with each other, the error of FBG#2 is just the opposite of FBG#1. As we can see in Fig. 4(b) and 4(c), these errors are nearly solved after adding the Hanning window. Therefore, by introducing the Chirp-Z + Hanning window algorithm, the peak-finding accurate of multi-grating is effectively improved even with only one cycle of time domain signals, which makes the high speed demodulation easily realized. For the zooming resolution of 50 Hz, the theoretical wavelength resolution of the system is about 0.625 pm according to Eq. (8).

We fabricated 105 identical weak FBGs with 1550 nm center wavelength along one optical fiber to verify the performance of the proposed system. The spacing of FBGs is 1 m, and the peak reflectivity is 0.1 %. The frequency bandwidth $B$ and the sweep cycle $T$ of MW signal are set as 1 GHz and 25 μs, and the sampling rate of DAQ is 2 GHz. The length and total dispersion of DCF are about 14.7 km and -2.03 ns/nm, respectively. In order to satisfy the constraints of Eqs. (4) and (5), a section of delay fiber (500 m) is attached to the coupler 2. With calculation, the frequency aliasing will not occur and high speed demodulation can be realized.

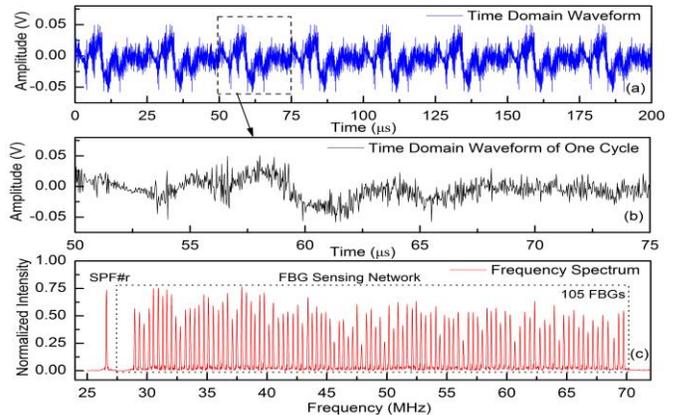

Fig. 5. The waveforms of 105 weak FBGs (a) time domain waveform of eight cycles, (b) one cycle of time domain waveform, (c) frequency spectrum waveform.

Fig. 5(a) is the time domain waveforms acquired by the DAQ. One cycle of time domain signals are intercepted from Fig. 5(a), as seen in Fig. 5(b), and processed by the Chirp-Z + Hanning window algorithm, so the frequency spectrum can be obtained, as Fig. 5(c) shown. In Fig. 5(c), each grating can be distinguished well with the demodulation of this system. The reason why the amplitude of the spectrum for each grating fluctuates greatly is that all of FBGs are written by hand. In the following, we tested the demodulation system by temperature, strain and vibration experiments.

In long time demodulation, the length disturbance of DCF will greatly affect the change of difference frequency signal, resulting in serious error. We eliminate it by adopting the single-peak filter as a reference. Temperature with DCF is carried out to verify it. During the experiment, DCF is put into the thermotank whose temperature changes from 6 to 51 ℃, and FBG#1 of the sensing network is stick on the thermostat with a constant temperature.

The thermotank is used to simulate the external disturbance, and we collect data every 3 ℃. According to Eq. (8), demodulations are performed in two ways, one is based on the frequency change of FBG#1, and another is the use of reference. The results are shown in Fig. 6(b), dotted line and solid line, respectively.

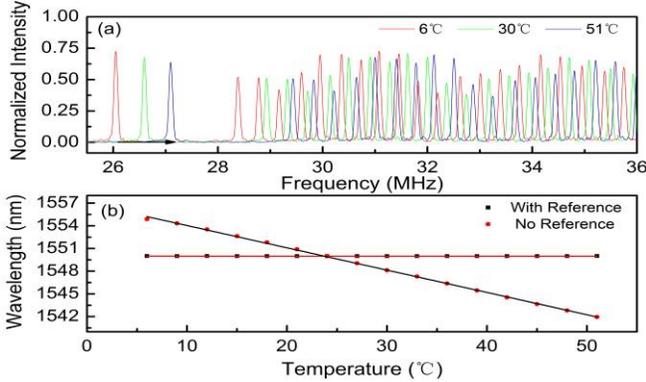

Fig. 6. The results of temperature experiment with DCF (a) the change of frequency spectrum, (b) the demodulation results of two ways.

In Fig. 6(a), the frequency spectrum shifts as a whole when DCF is heated. The difference frequency of FBG#1 changes, even if the wavelength of FBG#1 is constant. Thus, if based on the frequency change of FBG#1, the demodulation error will very larger. As illustrated in Fig. 6(b), the max change and standard deviation (STD) of demodulation results are 12930 pm and 4211 pm, respectively. However, by using the single-peak filter as a reference, the demodulation results almost unchanged. The max change and STD of results are 8 pm and 2.1 pm. Consequently, the using of single-peak filter as a reference, can effectively weaken the influence of external disturbance, which greatly improved the stability and accuracy in long time demodulation.

Next, the feasibility and accuracy of the system are verified by the strain experiment. In order to compare the demodulation results with the Optical Spectrum Analyzer (OSA), at the port 3 of the circulator, a part of signal is transmitted to the OSA by a light splitter. Since OSA cannot demodulate identical FBGs array, a large bend loss is created between the FBG#1 and FBG#2, so that only the reflected light from FBG#1 can be returned. Before the strain experiment, the frequency difference of FBG#1 and the single-peak filter is calibrated to be 2.329 MHz. Taking into account the mechanical strength of FBG, the strain varies in the range of 0 με to 1500 με, and the interval is 50 με. During the strain applied to FBG#1 with a precision adjust device, we measured the difference frequency signal by DAQ.

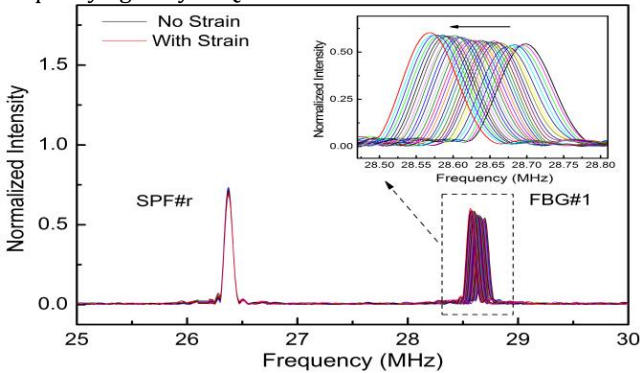

Fig. 7. The transformation of grating spectrum to frequency domain.

After processing by the Chirp-Z + Hanning window algorithm, Fig. 7 is the frequency spectrum. In Fig. 7, the frequency spectrum shift when FBG#1 is subjected to strain. Due to DCF with a negative dispersion coefficient, when the center wavelength of FBG in the OSA shifted to right for 0.1054 nm, the frequency spectrum shifted to left for 8400Hz. So the wavelength variation can be calculated for 0.1034 nm by Eq. (8), which is basically the same with the OSA. Thus it is proved the proposed system is feasible, the conversion of frequency to wavelength can be achieved well.

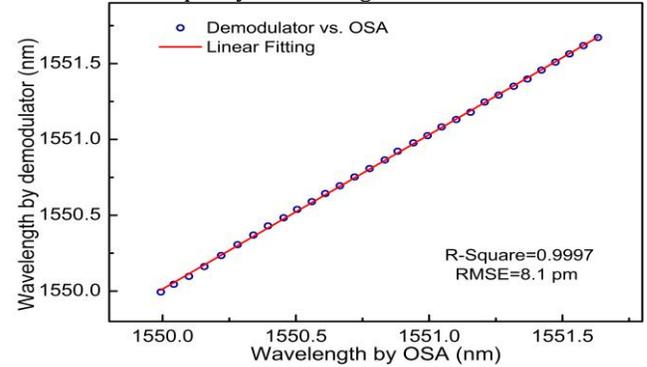

Fig. 8. The demodulation results of the strain experiment.

The results of the system and the OSA at same strain condition are also shown in Fig. 8, where X axis represents the wavelength measured by the OSA and Y for this system. In Fig. 8, linear fitting of measuring points are made. The R-square of fitting curve is 0.9997, and the Root mean square error (RMSE) is 8.1 pm. Which means the demodulation accuracy of this system is about 8.1 pm.

Lastly, we carried out a high speed vibration experiment to verify the 40 kHz demodulation speed of the system. Specifically, the last two weak FBGs of the sensing network are made as vibration sensors and fixed on the vibration test bench (Denmark B&K Company). With the action of 2 kHz sine vibration signal, the system has demodulated the wavelength of FBG sensors.

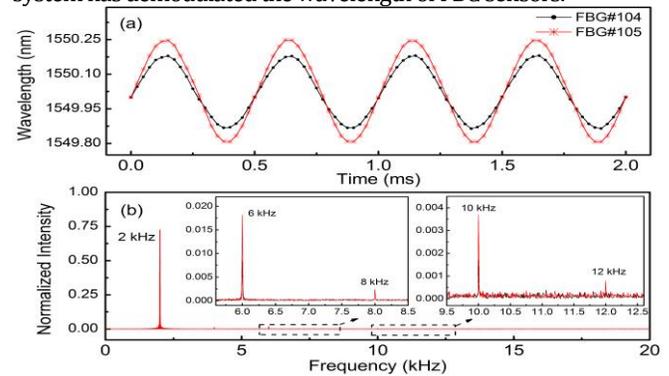

Fig. 9. The demodulation results of dynamic signal (a) wavelength domain diagram, (b) the frequency spectrum distribution.

Fig. 9(a) shows the wavelength varies in the time domain, in which black line represents the demodulation results of FBG#104 and red line for FBG#105. In Fig. 9(a), there are 40 sampling points in two cycle of 2 kHz vibration signal, and the time interval of each sampling points is 25 μs. Similarly, another FBG is the same condition, and the difference in the change of wavelength is due to the sensitivity of the two vibration sensors. According to the sweep cycle $T$ and the sampling points of vibration signal, it can be analyzed that the demodulation rate of the system reaches 40 kHz.

Fig. 9(b) shows the frequency spectrum distribution acquired by making FFT of the vibration wavelengths. As we can see in Fig. 9(b), the vibration frequencies of two sensors demodulated by the system are 2 kHz, which are consistent with the frequency we have set. In the inset graph, the high frequency doubling of 6, 8, 10 and 12 kHz can be seen clearly. Accordingly, the vibration experiment demonstrate that the system is ideal for the demodulation of vibration signals, with complete waveform and high response, and the rate is up to 40 kHz.

In summary, we propose and realize a high speed quasi-distributed demodulation system for weak FBGs based on the microwave photonics and the chromatic dispersion. Due to the dispersion effect of DCF, wavelength shift of FBGs are transformed into frequency domain. With the way of crossing microwave sweep cycle, all wavelengths of cascade FBGs can be high speed demodulated by measuring the change of difference frequency signal. Moreover, through adopting the Chirp-Z and Hanning window algorithm, the analysis of difference frequency signal which demodulation based is achieved very well. And by using the single-peak filter as a reference, the length disturbance of DCF caused by temperature is also eliminated, which makes the accuracy and stability further improved. Compared to traditional FBG wavelength demodulation technologies, this novel method provides many advantages such as higher speed and lower sampling rate. Experiments show that the system can realize the wavelength demodulation of weak FBGs with 0.1% reflectivity, 1 m spatial interval. It has a good static and dynamic demodulation capability, while the demodulation rate is as high as 40 kHz, the accuracy is about 8 pm, respectively.

**Funding.** National Natural Science Foundation of China (61575149).